\documentclass{egpubl}
\usepackage{eurovis2020}

\SpecialIssuePaper         

\usepackage[T1]{fontenc}
\usepackage{dfadobe}

\usepackage{cite}  
\BibtexOrBiblatex
\electronicVersion
\PrintedOrElectronic
\ifpdf \usepackage[pdftex]{graphicx} \pdfcompresslevel=9
\else \usepackage[dvips]{graphicx} \fi

\usepackage{makecell}
\usepackage{subcaption}
\usepackage{egweblnk}

\newcommand{\etal}{\textit{et al.}}

\usepackage[normalem]{ulem} 
\usepackage{xcolor}

\title[EuroVis 2020]%
      {SirenLess: reveal the intention behind news}

\author{Xumeng Chen, Leo Yu-Ho Lo, Huamin Qu \\ Hong Kong University of Science and Technology}

\begin{document}

\teaser{
 \includegraphics[clip, trim=0cm 4.38cm 0cm 0cm, width=\linewidth]{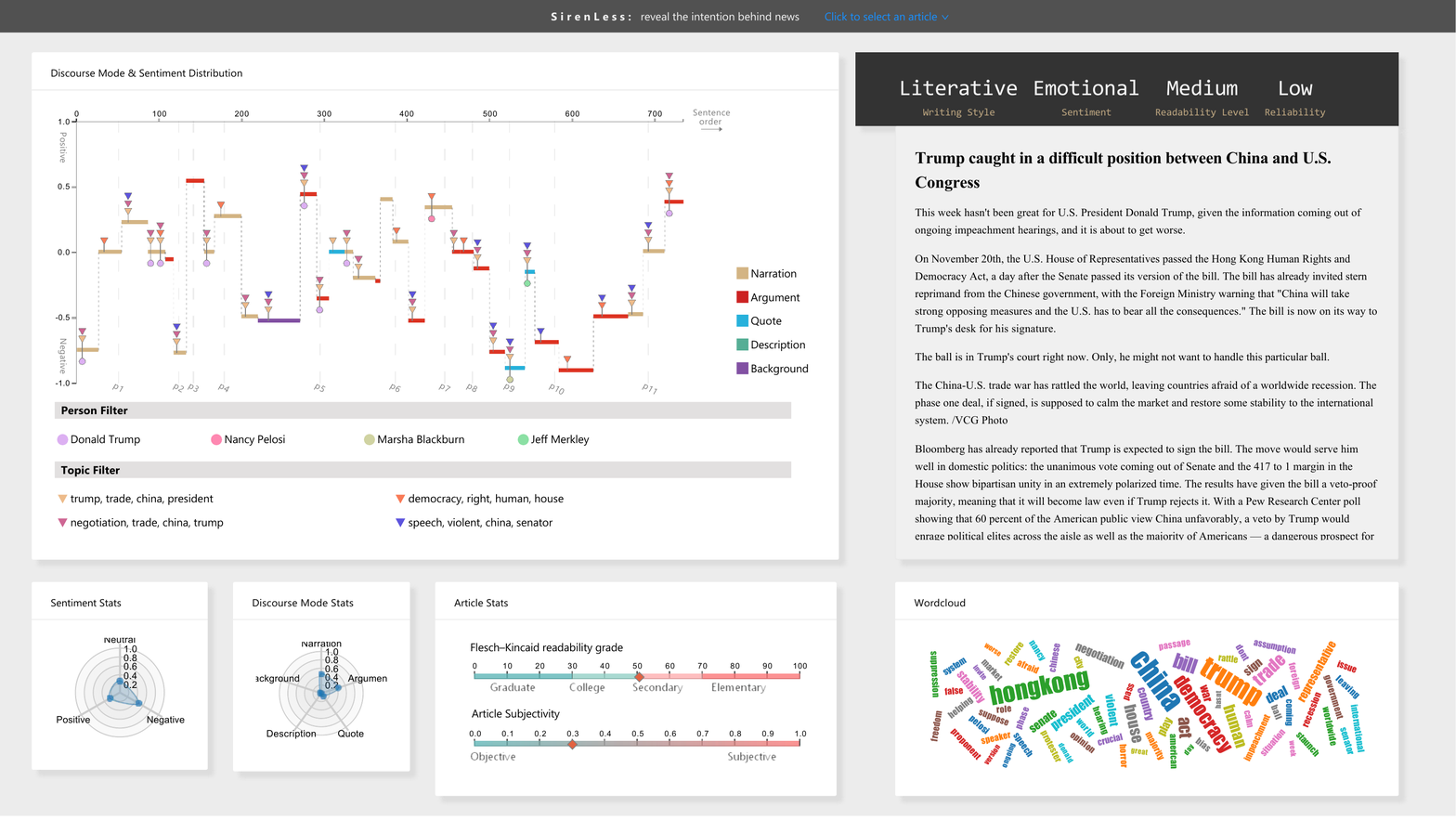}
 \centering
  \caption{A snapshot of SirenLess. At the top-left is the Article Explorer, the main pane of the system. }
\label{fig:teaser}
}

\maketitle
\begin{abstract}
News articles tend to be increasingly misleading nowadays, preventing readers from making subjective judgments towards certain events. While some machine learning approaches have been proposed to detect misleading news, most of them are black boxes that provide limited help for humans in decision making. In this paper, we present SirenLess, a visual analytical system for misleading news detection by linguistic features. The system features article explorer, a novel interactive tool that integrates news metadata and linguistic features to reveal semantic structures of news articles and facilitate textual analysis. We use SirenLess to analyze 18 news articles from different sources and summarize some helpful patterns for misleading news detection. A user study with journalism professionals and university students is conducted to confirm the usefulness and effectiveness of our system.


\printccsdesc   
\end{abstract}  
\section{Introduction}

With the commercialization and popularization of journalism, the authority and credibility of news articles are now being questioned. It is not rare to see writers with ulterior intentions use certain tricks to entice their readers to think in a certain way \cite{mcmanus2009commercialization}. However, it can be a difficult task for untrained readers to identify those latent intentions and stay alerted and objective while reading. Although heated discussions have occurred in academia concerning "fake news", "misinformation" and "click-baiting" \cite{bakir2018fake, aisch2016dissecting, vosoughi2018spread}, there has not been any universal solution to the problem, and moreover, few visualization work has contributed to this field. 

Much pioneer work has been carried out by NLP researchers on detecting fake news and misleading information.  Popular solutions including TF-IDF using bi-gram frequency \cite{li2007keyword}, probabilistic context free gram-mars, or PCFGs) \cite{feng2012syntactic}, and a combined feature union \cite{conroy2015automatic}, mostly focus on automatic feature extraction and classification. While useful for massive information filtering, the high cost of manual labeling limits analyses to small samples. Also, owing to their black-box nature, machine learning approaches are deficient in human-readable evidence to support their judgment, thus provides limited help for people when it comes to individual cases. 

The visualization approach can be a satisfactory solution to aid misinformation detection because it is human-friendly and self-evident. Nevertheless, because of the complexity of this problem and the lack of computational tools, few solutions have been proposed to address it visually, making it a no-man's land in visualization research. 

In this paper, we present SirenLess, a news veracity assessment system focusing on linguistic cues. SirenLess automatically extract multiple linguistic features from a news article and encode them into visualization channels. In addition, it provides critical information such as news characters and topics and show their distribution in the article. The linguistic features are elaborately selected based on domain-specific analytical methods and state-of-art NLP techniques. We validate our methodology and SirenLess via an informal user study. 

In summary, we contribute: (1) an integrative visualization design that could act as an entry point or hint for further visualization research on information veracity assessment. (2) A field study on current computer-based news analytical techniques. (3) A case study reporting the patterns found by our methodology, and a user study demonstrating the effectiveness of this system. We plan to make SirenLess available as an open source resource for people's use, together with all samples cases in this paper for easy replication.

\section{Related Work}

Combating fake news and misinformation is a high priority work in every civilized society. Lazer \etal have written a good introduction to the nature and impact of fake news \cite{lazer2018science}. The article suggests one technological solution that can help mediate the negative effect of fake news is to provide assistance tools to help individuals evaluate fake news.

One particularly dangerous kind of fake news is one intended to provoke emotions towards the news story \cite{bakir2018fake} and lead to extremist behavior \cite{aisch2016dissecting}. In addressing this kind of malicious intent, rhetorical structure and discourse analysis are promising directions \cite{conroy2015automatic}. The articles written with intent to have a distinguishable difference in rhetorical structure to truthful news \cite{rubin2015truth}. To bring readers into action, this kind of articles emotionally targets the readers and is termed ‘empathic media’ \cite{bakir2018fake}. Outstandingly, the disinformation on vaccination is very effective on despairing parents who seek an explanation on the cause of their child's misfortune.

Significant efforts have been invested in analyzing the spread of fake news, trying to capture the pattern of the spread of misinformation and the factors contributing to it \cite{vosoughi2018spread}. Through analyzing social media platforms, fake news is well crafted to be novel, so that people are more likely to share or reply, further helping the spread of falsehood. However, once debunked, the spread is quickly halted.

Instead of looking into the transmission of fake news, analyzing the content of a news article itself can flag the malicious intent of the writer. Some experimental attempts on automated deception detection show promising results. Through syntactic stylometry analysis, Feng et al. demonstrated 80\% to 90\% accuracy on different product review datasets \cite{feng2012syntactic}. In addition, the study conducted by Rubin and Lukoianova shows that deceptive stories are separable from truthful stories in the rhetorical structure feature space.

An alternative to automated detection of fake news is to equip human readers with the ability to evaluate the credibility of any fake news they encounter. Fact-checking websites such as FactCheck.org \cite{factcheck2019} and PolitiFact \cite{politifact2019} verify claims made on news articles, while Ad Fontes Media rates different news sources based on their reliability and bias \cite{mediabias2019}. Going further, education is the most important and effective way to prevent any damage done by misinformation, a course offered by University of Washington is a very good example \cite{bergstrom2017calling}.

This work resonates with the call for spotting disinformation from questionable news articles and by presenting the readers a stylometry analysis to debunk the polarization intent of the news writers.

\section{Design Goals}

The overarching goal of our work is to investigate any significant linguistic features of news articles to assist veracity assessment. 

Our design was initially inspired by real-life experience and demands, which is further consolidated by Conroy et al.'s survey on automatic deception detection \cite{conroy2015automatic}. To maximize the delivery of linguistic features out of news articles and at the same time, make the visual channels self-evident to common users, we formulate the following design goals.

\textbf{G1:  Provide a quick overview of the language usage of news articles.}  According to the theories of fake news detection, subjective writers use their languages strategically to avoid being caught \cite{feng2013detecting}. Through revealing the language usage of news articles, we vastly ease the procedure of finding out possible language leakages of untrustworthy news articles for common readers. We instantiated the theory by dividing it into four visualization tasks: visualizing discourse modes, sentiment, subjectivity and readability level. (T1, T2, T3, T4)

\textbf{G2: Present news meta data to help users grasp its semantic structure.} The language usage information is considered to be secondary information, where plain text has been processed by NLP algorithms to extract these information. To help readers better grasp the semantic structure and writing strategies of news stories, additional meta data such as characters and keywords, should also be delivered to users as important reference \cite{bal2009narratology}. 

\textbf{G3: Let users gain direct access and reference to the article text.} Studies have shown that a close reading of a script can enable deeper analysis of the context information \cite{janicke2015close, mccurdy2015poemage, kim2017visualizing}. While a high-level, abstract visual summary can be useful in analyzing news articles, in the end, it is no more than an auxiliary means, the final comments should still be made by a reader based on the raw text. 

Based on these design goals and currently available NLP techniques \cite{loria2014textblob, honnibal2017spacy, song2017discourse}, we derived four main visualization tasks listed below.

\textbf{T1: Reveal the sentiment and discourse mode distribution of the article.} We visualize the sentiment and discourse mode associated with each sentence. An aggregated view is also provided for quick reference. [G1]

\textbf{T2: Identify the estimated subjectivity and readability level of the article.} We chose the Flesch reading-ease test \cite{flesch2007flesch} to evaluate the readability level in this task. [G1]

\textbf{T3: Identify and compare character and keywords occurrences in the article} and allow users to select/ deselect them. [G2]

\textbf{T4: Provide the original text} and allow users to associate sentences with visualization entities. [G3]

We further elaborate the implementation of these tasks in the system design section.

\section{System and Data}

\subsection{Data Processing}

To make SirenLess a handy tool for end users, we established an automatic data processing pipeline to extract high-level semantics from plain text, which covers discourse mode, sentiment, readability, subjectivity, characters, and keywords analysis. On average, it takes 15-20 seconds for the pipeline to generate all required data from the raw text, while this figure may vary as the length of article varies.

\textbf{Discourse mode analysis.} In general text analysis, rhetorical modes are categorized into four types: narration, description, exposition and argument. This taxonomy is suitable for general purposes, but when the scope is narrowed down to news articles, a modified one can do better in describing the property of a sentence. Taking both general taxonomy and Tom Wolfe's theory \cite{wolfe1972birth} into consideration, we decided to classify news discourse mode into five categories: (1) Narration: the most important part of storytelling, which is the author's interpretation of the story; (2) Argument: analysis and ideas of the author; (3) Quote: directly repeat the passage of a person; (4) Description: detailed depiction, aimed at rebuilding the original scene; and (5) Background: fact-checked background information aimed at helping readers understand the current story. 

We finished the sentence classification task using the state-of-the-art method \cite{song2017discourse}, which adopted a multi-label neural sequence labeling model for discourse mode identification. While our discourse mode taxonomy is slightly different from that in this paper, since the features to be learned by the neural network  remain the same, we simply modify the output layer of the neural network, increasing the number of neurons from four to five. The model has an average F1-score of 0.7.

\textbf{Analysis of other semantics.} We employed gensim \cite{vrehuuvrek2011gensim}, NLTK \cite{loper2002nltk}, and SpaCy \cite{loper2002nltk}, which are considered to be popular and robust APIs in natural language processing, to extract other semantics from the plain text. We adopted LDA model \cite{loper2002nltk} to get possible topic descriptions of the article, where topics are represented as a sequence of keywords.

\subsection{Training Data}

We collected 312 news articles from multiple channels, including Fox News, ABC news, New York Times, the Economist and so on, to minimize the bias of training data. With reference to work from the Pew research center \cite{mitchell2014political}, we make a rough pre-filtering of articles to ensure the relatively even distribution of polarized and unpolarized news articles.


\section{Visual Design}

With the intention to make the interface self-evident and user-friendly, the final design is composed of three functional modules, each of which consists of several components. In this section, we will introduce the visual representations and user-interaction functionalities in detail, and discuss our considerations in making these design decisions.

\subsection{Article Explorer Module}

Article explorer (Fig 1, top-left) , which visualizes the distribution of discourse modes, sentiment and metadata in the news article, is the core of SirenLess. The integration of these linguistic features can evolve to high-level information, say critical sentences of the article, emotional tendency of the writer, and the like. A similar curve could be seen in Nam et al.'s work to visualize nonlinear narratives in movies \cite{kim2017visualizing}, while the use of visual channels are completely different, and we also label the occurrence of characters and topics in a novel way. The article explorer has three main features illustrated as follows.

\textbf{Revealing Sentiment Distribution}
When analyzing the sentiment score of news articles, it is a notable problem that after being averaged, the sentiment score of articles tend to be close to zero and indistinguishable, which increases the difficulty of utilizing the sentiment as an indicator to detect misinformation. To handle this, we propose to visualize the distribution of sentiment of the whole article, where sentimental outliers could easily be perceived and call on the vigilance of users. We also provide an overall sentiment score in the article stats block.

Each sentence's corresponding sentiment score is associated with a float number between -1 and 1 to distinguish negative ones from positive. Through testing, we found sentences with extreme sentiment whose scores of absolute value greater than 0.5 could easily draw user's attention. It is worth mentioning that through associating sentiment with metadata(character and topics), users could gain insights into the standpoint of the article, i.e. the author's emotional tendency towards certain characters and topics, which is of great help in analyzing the article.

During our design process, we also considered a vertical layout for the panel to observe the metaphor of the article reading order from top to bottom. However, this design was voted down right away when the draft came out. The shape of computer screen restricts the space for a vertical layout, and a readable graph calls for scrolling, which outweigh the flaw of giving up the metaphor.

\textbf{Revealing Discourse Mode Distribution}
We choose color as the visual channel to identify different discourse modes. We carefully design the colorway so it conforms to color psychology and deliver proper implication. For the most common discourse mode narration, a low-saturation, skin-like color is assigned to help emphasize other colors. A noticeable reddish color is assigned to argument, for they tend to be subjective and suspicious. Purple is assigned to background information as purple represents wisdom. Finally, we choose green and sky-blue, two positive colors to encode description and quotes since the two discourse modes are considered to be safe and reliable. 

\textbf{Reveal Metadata of News Articles}
To better analyze an author's writing strategy, it is important to visualize any additional news metadata along the curve. In selecting metadata, we are inspired by the "5W\&1H" theory which formulated "who, what, where, when, why and how" to be the six principal elements in journalistic writing \cite{shabir2015enhancing}. Among the six elements, "why" and "how" are considered to be high-level knowledge that can not be extract properly by computer. Marking the position of the appearance of "where" and "when" in the article gives limited help to the understanding of news, and at the same time, the overuse of markers will cause confusion to users. Therefore, we decided to visualize "who" and "what" on the article explorer, where "who" is specified as characters, and "what" is specified as keywords grouped by topics generated by the LDA model \cite{ramage2009labeled}. Users could visually understand how the writer arranges the appearance of news characters and key information strategically to attain his potential goals. Also, since we visualized the splitting of paragraphs, it is not hard to gain an overview of how metadata are distributed in difference paragraphs.

When designing the glyph of the markers, we choose circles to represent characters, with the metaphor of human head, and triangles to represent keywords. We use colors with diverse hues to identify different characters and topic groups. Moreover, to visualize the co-occurrence of characters and topics in the same sentence, the markers are designed to stack up vertically in the same sentence (Figure 2). Users could easily figure out critical sentences in the article by looking at the height of the stack since higher stack means higher information density. This enables users to allocate their time wisely when reading the article. We also make filters for the markers to allow users to concentrated on data of interest, which also alleviates any possible problems of overwhelming visual channels.

\begin{figure}[htb]
  \centering
  \includegraphics[clip, trim=4.85cm 14.1804cm 21.8926cm 4.7cm, width=.8\linewidth]{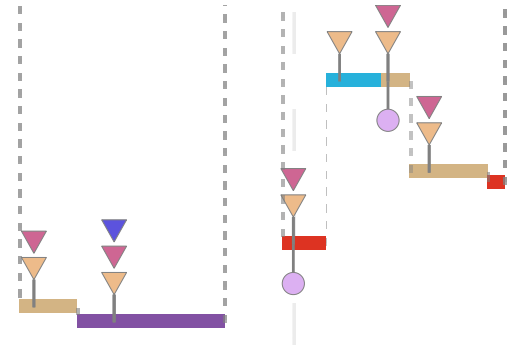}
  \caption{\label{fig:firstExample}
           Stacked markers}
\end{figure}

\subsection{Article Stats Module}
The article stats block aggregates highly summarized information for news articles, which helps users gain an overall insight into the article.

\textbf{Sentiment and discourse mode stats.} (Figure 3) The two radar charts are designed to work together with the article explorer. The latter could deliver sentence level information with the cost of relative weakness in revealing the overall situation and the previous one covers this perfectly. The stats view is also convenient for users who do not have the patience to look into the article explorer.

\begin{figure}[htb]
  \centering
  \includegraphics[clip, trim=0.5cm 5.0804cm 21.1926cm 11.7cm, width=.8\linewidth]{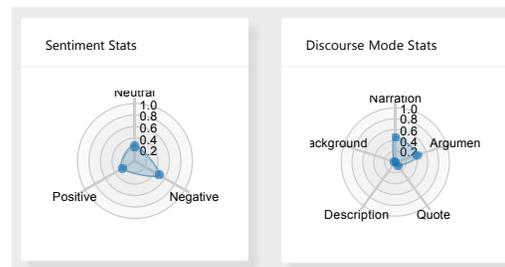}
  \caption{\label{fig:firstExample}
           Radar charts}
\end{figure}

\textbf{Article stats} (Fig 4)  which includes Flesh-Kincaid readability grade and article subjectivity. The readability grade, proposed by Flesh et al. \cite{flesch2007flesch}, reveals the difficulty level of an article by its vocabulary use. We include it as an index of detection since misleading news likes to be easy-to-read so it can spread faster throughout the general public \cite{conroy2015automatic}. The article subjectivity is an experimental feature generated by cutting-edge NLP technology, which can serve as a reference to users.

\begin{figure}[htb]
  \centering
  \includegraphics[clip, trim=8.75cm 4.5804cm 12.4926cm 11.7cm, width=.8\linewidth]{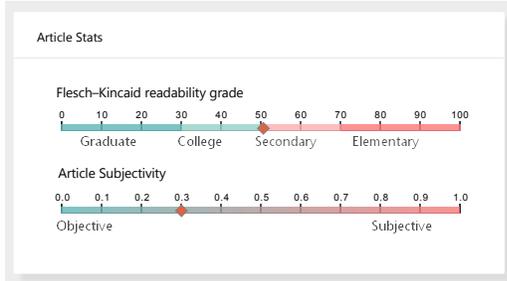}
  \caption{\label{fig:firstExample}
           Article stats}
\end{figure}

\subsection{Reader View Module}

In addition to the relatively sophisticated visual panel on the left, the reader module is designed to take the right half of the interface, which aims to offer the most convenience to users. The reader module comprises a text-summary header, the original text of the news article and a word cloud.

\begin{table*}[t]
\centering
\caption{Calculation scheme of text summaries}
\label{table:summaryheader}
\begin{tabular}{| l | l | p{30em} |}
\hline
\thead{Attributes} & \thead{Levels}                 & \thead{Scoring}                                         \\ \hline
Writing Style      & Rigorous, Balances, Literative & \makecell[l]{Grade: $( \#argument + \#quote - \#description - \#background ) / \#narration$ \\ Grade: $[0, 0.2), [0.2, 0.4), [0.4, 1]$} \\ \hline
Sentiment          & Calm, Regular, Emotional       & Article Polarity: $[0, 0.1), [0.1, 0.2), [0.2, 1]$ \\ \hline
Readability Level  & Easy, Medium, Hard             & Readability Score: $[0, 30), [30, 70), [70, 100]$  \\ \hline
Reliability        & Low, Medium, High              & \makecell[l]{Grade: $ 100 - (polarity + subjectivity)*100 $ \\ Grade: $[0,40), [40, 70), [70, 100]$} \\ \hline
\end{tabular}
\end{table*}

\textbf{Text-summary Header.} (Figure 5) This part did not appear in our original design, while it was added later during our testing of the interface. It is specially designed for first-time users and inpatient users. First, it reduces the difficulty for first time users to understand the usage of the whole system since it serves as hint for them to understand the visualization. Second, such high level summary could give impatient users the most critical information they need. Participants in our user study have shown special interest in this part, especially for those without visualization background. The categorical summary text of an article is decided based on the scheme specified in Table 1.


\textbf{Original Text.} (Figure 5) To enable quick bidirectional reference between visual entities and the original text, we do not merely embed the plain text into the interface, but add selection and click events to them. When a user selects a sentence segment in the article explorer, its corresponding text will be highlighted on the right.  This is especially beneficial for detailed analysis of articles.

\begin{figure}[htb]
  \centering
  \includegraphics[clip, trim=17.3cm 13.9804cm 0.9926cm 1cm, width=.8\linewidth]{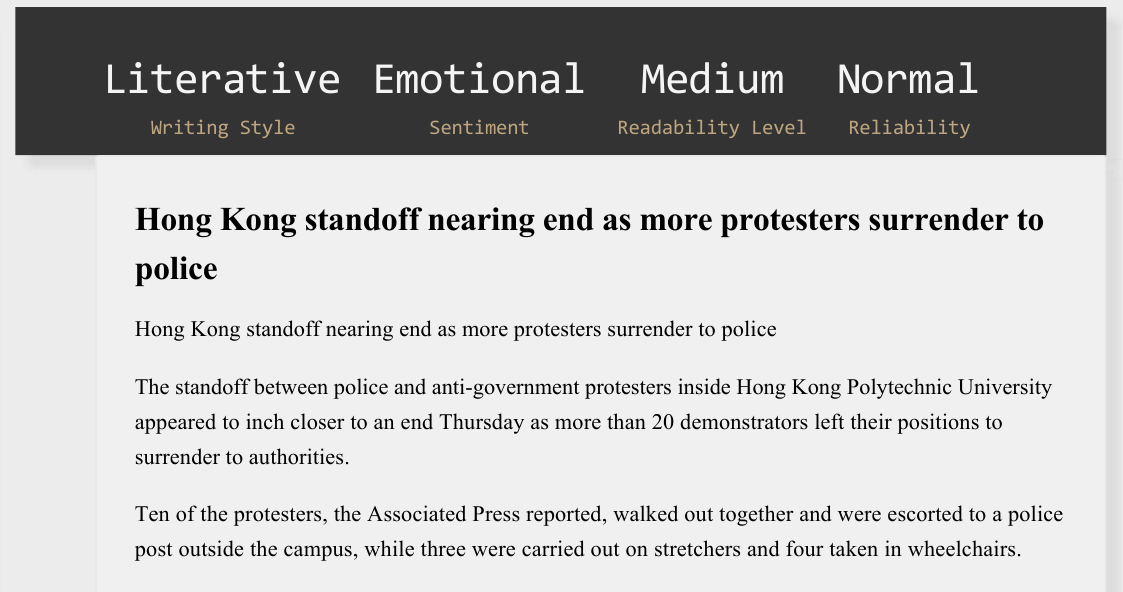}
  \caption{\label{fig:firstExample}
           Text summary header and Article viewer}
\end{figure}

\textbf{Word Cloud.} (Figure 6) Word cloud, as one of the most popular visualization tools at present, is known to be interesting and user-friendly, especially welcomed by users without visualization background. We provide a word cloud visualizing stemmed high-frequency words from the article to serve users' needs.

\begin{figure}[htb]
  \centering
  \includegraphics[clip, trim=18.1cm 4.5804cm 0.9926cm 11.8cm, width=.8\linewidth]{figures/article2.pdf}
  \caption{\label{fig:firstExample}
           Word cloud}
\end{figure}






\section{Patterns of unreliable news}

\begin{figure*}
    \centering
    \begin{subfigure}[b]{.5\textwidth}
        \centering
        \includegraphics[clip, trim=1cm 12.8804cm 14.6926cm 2cm, width=\linewidth]{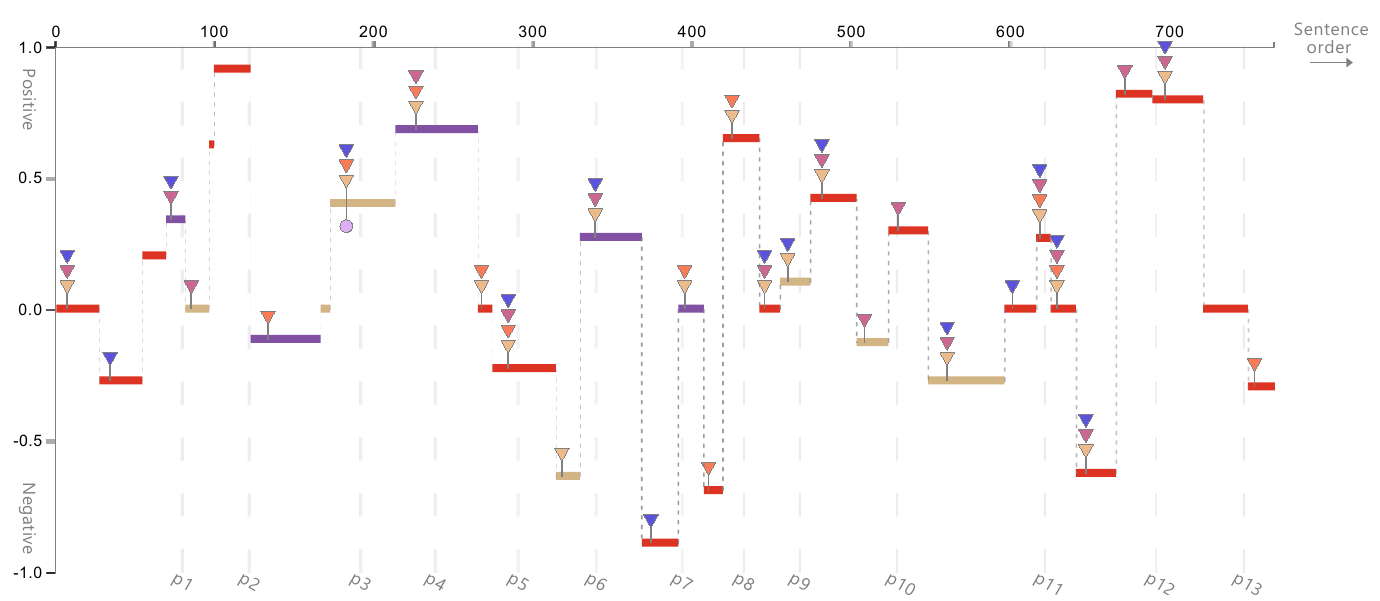}
        \caption{Jobs for life are a thing of the past.}
        \label{fig:sfig1}
    \end{subfigure}%
    \begin{subfigure}[b]{.5\textwidth}
        \centering
        \includegraphics[clip, trim=1cm 12.8804cm 14.6926cm 2cm, width=\linewidth]{figures/article2.pdf}
        \caption{Hong Kong standoff nearing end as more protesters surrender to police.}
        \label{fig:sfig1}
    \end{subfigure}
    \begin{subfigure}[b]{.5\textwidth}
        \centering
        \includegraphics[clip, trim=1cm 12.8804cm 14.6926cm 2cm, width=\linewidth]{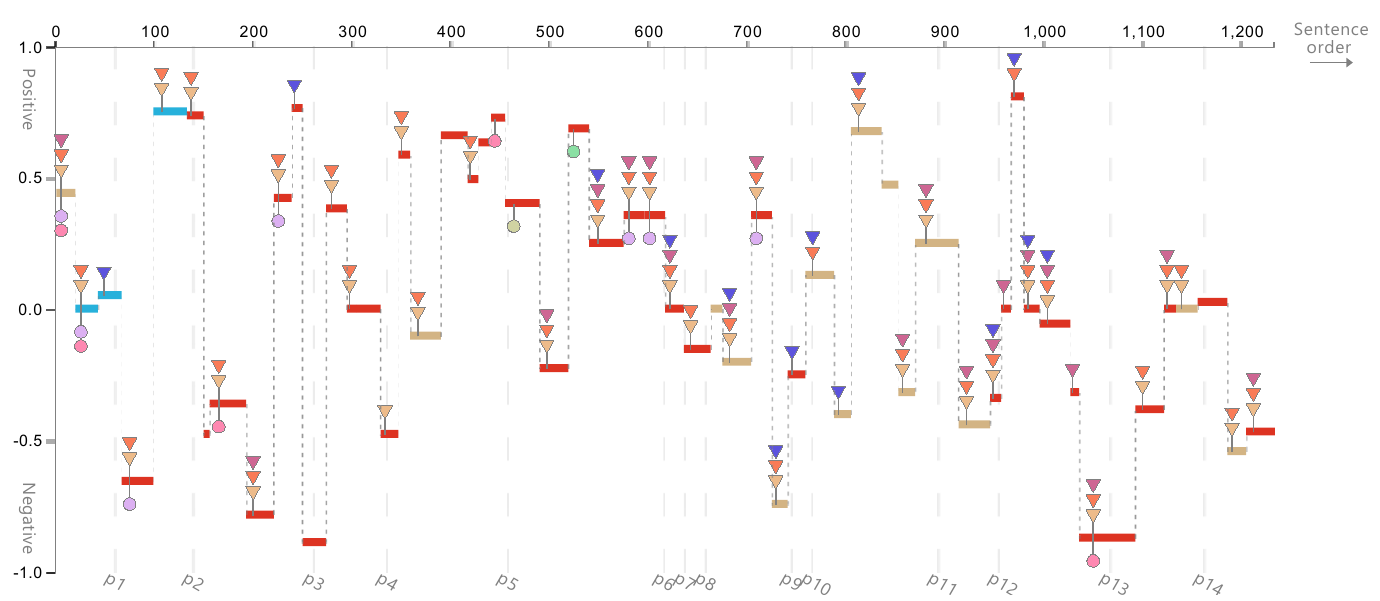}
        \caption{Donald Trump’s America Can’t Fight Xi Jinping’s China.}
        \label{fig:sfig1}
    \end{subfigure}%
    \begin{subfigure}[b]{.5\textwidth}
        \centering
        \includegraphics[clip, trim=1cm 12.8804cm 14.6926cm 2cm, width=\linewidth]{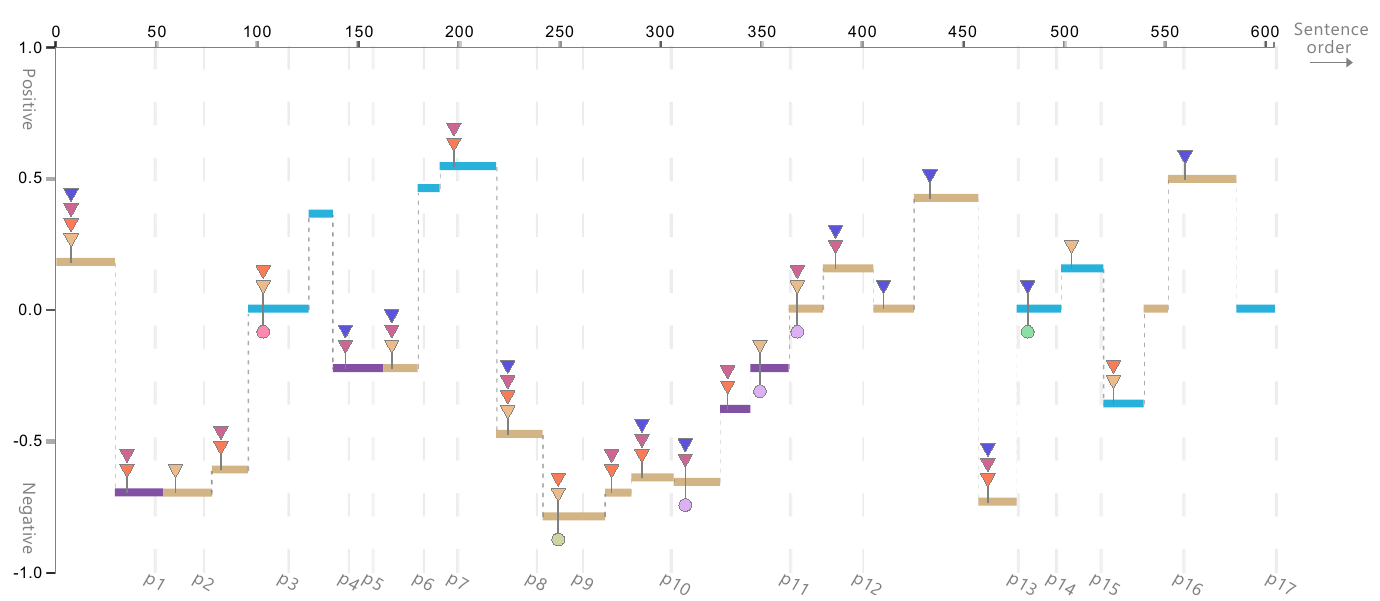}
        \caption{Hong Kong protesters take their cause to airport arrivals.}
        \label{fig:sfig1}
    \end{subfigure}
  \caption{\label{fig:articleexplorer}Article Explorer for serveral articles}
\end{figure*}

While one principle of our system design is to provide the most freedom for users to analyze news articles as the wish, we also provide some guiding visual patterns of unreliable news for users to choose from and build their own analyzing pipelines. To achieve this goal, we analyzed 8 misleading articles using our system. Results show that some linguistic features (sentiment distribution, article subjectivity, article readability) have strong indicative patterns on their own, and some work together to reveal certain patterns. We hereby summarize four patterns for singular features, and two for composite features.

\subsection{Singular features}
\textbf{Sentiment distribution.} Reliable news is supposed to be "calm", where the author keeps his/her position as a bystander \cite{maras2013objectivity}.  On the contrary, unreliable news tends to be emotional, which could exits in two forms: dominated by one polarity (Figure 7a, 7b, 7d), or fluctuating between two polarities (Figure 7c). Both can be commonly seen in news articles and clearly visualized by article explorer. For the first type, the writer uses most of sentences to narrates one side favorably or to give adverse comments. For the second type, the writer deliberately arrange two forces of confrontation, giving one side a positive slant, and the other negative.

\textbf{Article subjectivity.} Subjectivity is another strong indicator to show the reliability of news articles. Based on the algorithm we adopted, we empirically get the result that reliable articles could have a subjectivity score less than 0.2, while unreliable ones generally greater than that.

\textbf{Readability grade.} According to known theories in communication study, an article should be easy to understand if it wants to be popular \cite{conroy2015automatic}. Similarly, unreliable news shows an easy-to-read pattern with a Flesch-Kincaid readability grade greater than 0.3 in our study, which is approximately equivalent to the cognitive level of secondary and college level students. Reliable news's grades usually fall in the cognitive range of college and graduate levels.

\textbf{Discourse mode distribution.} While less obvious than the previous features, certain patterns exist in the discourse mode distribution. First, although not necessarily dominant, there are a considerable portion of arguments (Figure 7a, 7c), while reliable news should be narrative without subjective judgments. Second, although background information is considered to be a signal of reliable news, sly writers may arrange descriptions strategically with arguments, with the intention of distorting facts with their personal opinions (Figure 7a).   

\subsection{Composite features}

\textbf{Sentiment and characters.} Misleading news tend to have strong preference towards certain characters (Figure 7a, 7b, 7c, 7d), where characters are associated with emotional sentences. We can see a clear pattern by combining the person filter and sentiment distribution curve together. For example, in the article written by an U.S. writer talking about the relationship of U.S. and Chinese presidents concerning the issue of Hong Kong protesters (Figure 7c), the U.S. president is associated with positive sentiment while the Chinese president is associated with negative ones.

\textbf{Discourse mode and sentiment.} Quotes have the effect of making an article look more trustworthy since it seems that the ideas come from tertiary sources rather than the author. However, sophisticated writers selectively quote people's words, using others' mouth to convey their words. This phenomenon is easy to be ignored when reading plain text, while it can be seen clearly through visualization (Figure 9). Special attention should be called when there are a lot of quotes appearing in the article, with most of them associated with strong sentiment.

\begin{figure*}
    \centering
    \includegraphics[clip, trim=1cm 12.8804cm 14.6926cm 2cm, width=\linewidth]{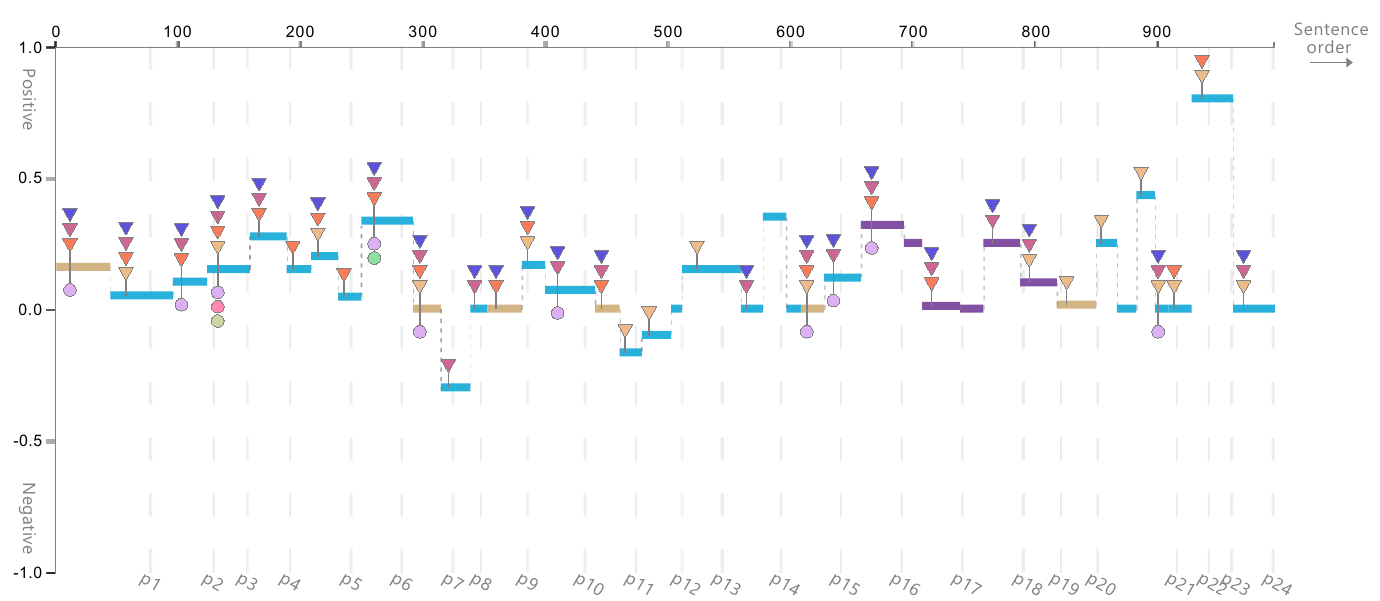}
    \caption{Super-rich prepare to leave UK 'within minutes' if Labour wins election}
    \label{fig:my_label}
\end{figure*}

\section{Evaluation}

\subsection{Journalism Scholar Review}

Because of time limit, we invited one journalism professor to evaluate SirenLess through a questionnaire. After using our system, the professor gave an overall comment to it: "Meaningful and potentially useful tool for news/information analytics". He described the analytical features shown on the dashboard to be "clear and informative" from an expert's point of view. He affirmed the usefulness of linguistic features we adopt in our system: "The features of writing style, sentiment, and keywords, etc. can be relevant indicators of journalistic performance, depending on the usage context."

The professor mainly commented three use cases of our system. The first one is for academic research in media content analysis for media research companies and scholars: "The news analysis provided by the system can help readers/researchers to obtain a better understanding of the linguistic features of particular news pieces." The second one is for teaching and learning in journalism and public relations courses. The third is for news room practice, especially for editors. 

\subsection{User Study}

We conducted a preliminary user study among college students to evaluate our system and verify whether it can help them to identify the misleading intention of the news articles. Our experiment invited 9 participants including 4 undergraduate and 5 postgraduate students. Among all the participants, 4 of them have formal visualization training, while others do not.

In the study, we first give a tutorial to help the user to understand the interface. We explain to the participants on different views of the visualization system with the use of two news articles, one is a propaganda-like article, which has strong sentiment, another one is more like a weather report, which contrasts with the propaganda one. Once the participants picked up with the interface, we start to conduct the experiment. The experiment is in two steps, first is read first and review with the visualization system, second is viewing the visualization first then read under the help of the visualization system.

In the first step, we first let the participants read a selected article without the help of the visualization system and tell if the article has a bias toward a certain entity. Since the source of the articles is a strong indicator of the participants' judgment, we deliberately conceal it during the experiment. After recording their answers, we let the participants review the article in the visualization system. Only 2 out of 9 had successfully spotted the bias of the article without the help of visualization, they found that one of the paragraphs has a strong misleading intention and violate the facts they knew. After reviewed the article with the visualization system, 8 out of 9 had successfully spotted the bias. The remaining one only agreed the article is biased after we revealed the source of the article.

In the second step, we let the participants view the visualization system of another two selected articles before reading it. The selected articles are concerning the same event but from two different newspapers and have significantly different standpoints towards the event. Same as the first step, we conceal the source of the article and record their answer regarding the article bias and misleading intention. First, participants can only interact with the visualization and are not allowed to read any of the article text. In these 18 samples, 7 out of 18 stated as highly alerted about the article bias, 5 out of 18 stated as dubious and 6 stated the articles look like objective in the visualization. The participants then allowed to read the article text while using the visualization system. When cross verifying the article text with visualizations, the participants previously stated dubious about the articles found actual evidence in the text and confirmed their doubt. The participants previously stated the article as unbiased realized the text is positive towards one side and negative towards the opposite side. After viewing both visualizations and article text, the participants unanimously agreed the articles are biased.

After the experiment, participants reply to a questionnaire regarding the usefulness of the visualization system. For the article as a whole, 4 participants replied strong agreement on the tools in helping them to spot the intention of misleading in the articles, 3 replied agreement and 2 replied as neutral. For spotting the bias towards different entities or events in the articles, all participants replied with an agreement and 5 replied with a strong agreement.

The results are very encouraging. As we work towards an easy-to-use system that common users can use and help the general public to realize the bias in the information they receive undoubtedly, we would test our system with a wider audience, especially with a higher variety in the education background.

\section{Discussion}

\subsection{Lessons learned from usability study}

The results of our user study indicate that our system has well accomplished our design goals. Also, the domain expert credit our work and see the potentiality of adopting it in research and teaching practice. But on the other hand, some limitations of our system are worth noticing.

\textbf{Accuracy of feature extraction.} Since our system highly relies on the accuracy of extracted data, although we have adopted the state-of-art method, there are some embarrassing scenarios where extracted features denies those suggested by human. This problem can be ameliorated by annotating a larger news corpus to improve the performance of NLP algorithms. Plus, it may be necessary to encode the uncertainty level and alert users of potential discrepancies.

\textbf{Interaction and readability.} Although there are many interactive designs in our system such as character/ topic filter and sentence highlighting, users still make further requests on the interactivity. We therefore outline two future work to improve the usability of our system: 1) Increase the functionality of the filter so the radar chart could change with it to reveal the related sentiment/ discourse mode distribution towards certain characters/ events. 2) Indicate the algorithm and calculation methods in layman terms. 

\subsection{Extending the system and supplement missing dimensions}

SirenLess mainly rely on linguistic features as indicators for journalistic performance.  While it is a good start for news analysis, it alone might not be sufficient for misleading news detection given that current outputs are more descriptive than conclusive. Suggested by domain experts, there are some possible directions for us to extend our system to better accomplish the goal of fake news detection.

The first one is to enable comparison by aggregating articles of the same topic. Currently, each news article is isolated and thus provides limited insights. We suspect it is necessary for the system to automatically collect news articles on the same topic from different resources and conduct batch analyses. This poses challenges on establishing an automatic data collection pipeline.

The second one is to integrate related external information such as comments and Wikipedia to enable fact check. Recent NLP researches have confirmed the viability of exploiting external resources \cite{pan2019improving, chen2017reading} for subject-area multiple-choice question answering. We may learn ideas from these NLP works and carefully devise visualizations for external information integration.

\subsection{Generalization to other domains}

Though the initial goal of the design of our system is news article analysis, we speculate that our system could be generalized to other genres, and would be especially useful for articles with clear pattern. One example is to help students analyze patterns of TOFEL writing samples. As a well-developed writing test with a huge sample pool, there are some patterns and templates for students to follow to get high grades. However, with English as their second language, it is hard for unexperienced students to grasp and understand those patterns. By providing sentence-level visual encoding of writing samples, students could quickly understand the writing style and linguistic features of sample articles. 


\section{Conclusion}

In this work, we present SirenLess, an interactive tool to detect misleading news articles by their linguistic features, including discourse mode, sentiment, subjectivity, readability level and so on. We illustrate several examples of our system and summarize some patterns of misleading news articles. The
findings accord with and further supplement the existing theories. We validate its usefulness through a user study and highlight its potential use cases in news research. In future work, we intend to extend the system by including more analysis dimensions. We also plan to evaluate the performance of our system on other text analysis tasks.

\bibliographystyle{eg-alpha-doi}
\bibliography{content/refs.bib}




\end{document}